# Optical spin waves

Vage Karakhanyan, Roland Salut, Miguel Suarez, Nicolas Martin, Thierry Grosjean*

FEMTO-ST Institute, CNRS, University of Franche-Comte, Besançon, 25030, France

*Corresponding author. Email: thierry.grosjean@univ-fcomte.fr

**Abstract:** Chirality is inherent to a broad range of systems, including in solid-state and wave physics. The precession (chiral motion) of electron spins in magnetic materials, forming spin waves, has various properties and many applications in magnetism and spintronics. We show that an optical analog of spin waves can be generated in arrays of plasmonic nanohelices. Such optical waves arise from the interaction between chiral helix eigenmodes carrying spin angular momentum. We demonstrate that these optical spin waves are reflected at the interface between successive domains of enantiomeric nanohelices, forming a heterochiral lattice, regardless of the wave propagation direction within the lattice. Optical spin waves may be applied in techniques involving photon spin, ranging from data processing and storage to quantum optics.

### 1- Introduction

Chirality is a universal symmetry property of matter and waves that is important in biology *(1)*, medicine *(2)*, physics *(3,4)*, and materials science *(5)*. In optics, chiral effects usually involve chiral light, characterized by helical waves carrying angular momentum *(6-8)*. The helical phase structure or field pattern rotation produces orbital angular momentum (OAM), while the rotation of the light polarization vector introduces spin angular momentum (SAM) *(9)*.

Analogies between electron and optical spin effects have recently led to the investigation of novel chiral states of light, such as photonic skyrmions *(10,11)*. Similar to their magnetic skyrmion counterparts, these spin-optical textures are topologically protected against external perturbations.

A wave-type manifestation of the electron spin is spin waves, which originate from spin precession in magnetic materials. These collective excitations in magnetically ordered media, quantized as magnons, result from the mutual coupling between precessing magnetic moments *(12)*. Spin waves propagate irrespective of the prevailing spin orientation and can be reflected at the boundary between ferromagnets with opposite magnetizations (i.e., developing spin precessions of opposite handedness) *(13-15)*. Similar to light, spin waves show polarization degree of freedom, which is locked to the circular eigenstate in ferromagnets, enabling spin precession with only one handedness *(16-18)*.

In this study, we show that an optical analog of spin waves in ferromagnets can be generated within a one-dimensional array of gold-coated carbon helices. These optical waves, which arise from near-field coupling between adjacent homochiral helices, possess an intrinsic angular momentum locked to the handedness of the nanostructures. These optical spin waves propagate irrespective of the helix direction and can be reflected at the interface between two domains of enantiomeric nanohelices, forming a heterochiral array, regardless of their propagation direction. The investigation of these optical spin waves may inspire new research on the manipulation of light, with direct applications in classical and quantum data processing and telecommunications. From a quantum perspective, an elementary optical spin wave excitation can be viewed as an optical magnon, a quasiparticle transported in a lattice of chiral nanostructures.

## 2- Spin properties of the constituent helical elements

In the traveling wave regime *(19)*, a metallic helical nanoantenna operates as a chiral optical waveguide that propagates twisted modes similar to spoof surface plasmons *(20-22)*. Right- and left-handed helices produce vortex eigenmodes with helical wavefronts with opposite intrinsic OAM. The intrinsic OAM is described by the vortex topological charge that governs the phase increment of the waves around the helix axis (i.e., the center of the vortex) *(23)*. The dispersion relations of the helix's vortex modes with angular momentum of L=±1 are discussed in Section S1 of the supplementary text *(24)*. In the considered spectral range, the helix has a plasmon-like guided mode and a leaky mode that show positive and negative effective refractive indices, respectively.

Owing to spin-orbit interaction *(9)*, these modes have longitudinal SAM aligned with the helix axis. The direction of this longitudinal SAM is determined by the handedness of the nanostructure and is either parallel or antiparallel to the propagation direction. In addition to the longitudinal SAM, the evanescent component of the helix eigenmodes generates transverse SAM oriented perpendicular to the mode wavevector. The interplay between these two components results in three-dimensional spin texture similar to that observed in optical Bloch-type skyrmion-like structures *(11)*. This spin distribution can be anticipated with the SAM density, which is defined as:

$$\boldsymbol{s} = \text{Im}[\epsilon \mathbf{E}^* \times \mathbf{E} + \mu_0 \mathbf{H}^* \times \mathbf{H}]/2\omega \tag{1}$$

where **E** and **H** are the electric and magnetic fields, $\omega$ is the angular frequency, and $\epsilon$ and $\mu_0$ are the permittivity and permeability, respectively *(25,26)*. The SAM density can also be written as:

$$\mathbf{s} = \frac{w}{\omega}\boldsymbol{\sigma} \tag{2}$$

where $w = 1/2\epsilon|\mathbf{E}|^2 + 1/2\mu_0|\mathbf{H}|^2$ is the energy density and $\boldsymbol{\sigma}$ is the local polarization helicity vector. The vector components of $\boldsymbol{\sigma}$ define the local helicity of the wave along the three spatial directions. The amplitudes of $\boldsymbol{\sigma}$ and its components are between 0 and 1 and -1 and 1, respectively. The values ±1 correspond to the circular polarization eigenstates. The SAM carried by the optical waves propagating along the helix is defined as $\mathbf{S}(z) = \iint \mathbf{s}(x, y, z)dxdy$, where (0, x, y) represents planes perpendicular to the helix axis (0z).

Figure S2 depicts the spin texture of the helix's guided mode. Fig. 1 shows that the total SAM of the traveling waves is aligned with the propagation direction. The orientation of the total SAM, either parallel or antiparallel to the wavevector of the guided mode, is defined by the handedness of the helix. Within the first turns of the helix, the total polarization helicity of the traveling waves is defined as:

$$\Sigma(z) = \frac{\omega}{W(z)}|\mathbf{S}(z)|, \tag{3}$$

where $W(z) = \iint \mathbf{w}(x, y, z) dx dy$ gradually increases up to 0.75 as the negatively refracted leaky mode decays within the structure (after a few turns, the helix becomes a monomode chiral waveguide). A total polarization helicity of 0.75 does not imply that the helix guided mode is elliptically polarized. Instead, this value reflects the contribution of the SAM to the total angular momentum of the guided wave owing to the spin-orbit interaction.

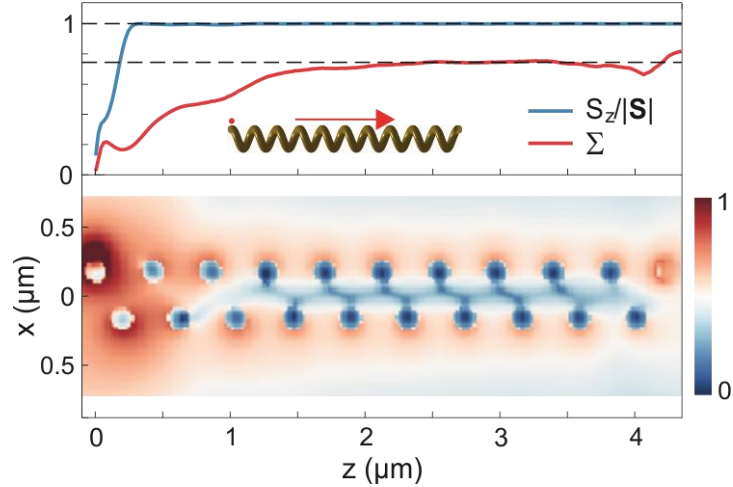

**Fig. 1. SAM orientation and polarization helicity of the twisted traveling waves of a plasmonic helix.** Bottom: Calculated field amplitude along a 10-turn helix excited by an x-polarized dipole positioned at its left end (see inset in the top figure). Top: Normalized projection of the total SAM ($S_z/|\mathbf{S}|$) and helicity of the twisted waves along the helix axis ($\Sigma$). Inset: Schematic of the simulation setup, where the red dot symbolizes the x-polarized dipole in contact with the structure, and the arrow indicates the propagation direction of the guided waves.

### 3- Response of the coupled helix array to dipolar excitation, optical spin waves.

When many helices are arranged in a one-dimensional array with a spacing that is smaller than the wavelength, the excited spoof surface plasmons are waves that propagate across the entire structure, as shown in Fig. 2 (**A**). In addition to energy transport along the helix axes, a coherent transfer of excitation between nanohelices occurs due to interactions between the helix-guided modes. Coherent energy transport via interparticle optical coupling has previously been demonstrated with arrays of closely spaced metal nanoparticles *(27-29)*. Near-field coupling between subwavelength waveguides can also produce metamaterials where optical waves freely propagate *(30)*. We can therefore generate bidimensional optical waves with angular momentum defined by only the intrinsic chiral properties of the support material.

The two key elements underlying such optical waves are the chiral eigenmode of the helix and the near-field interaction between the coupled helices. These two components can be viewed as optical analogs of the precession of the electron spin in the presence of an external magnetic field and the dipole-exchange interaction between atoms in a magnetic lattice *(12)*, respectively. In both regimes, the local spin effects are determined by the intrinsic chiral properties of matter, with geometrical chirality for optical spin waves and gyromagnetic properties for magnetic spin waves.

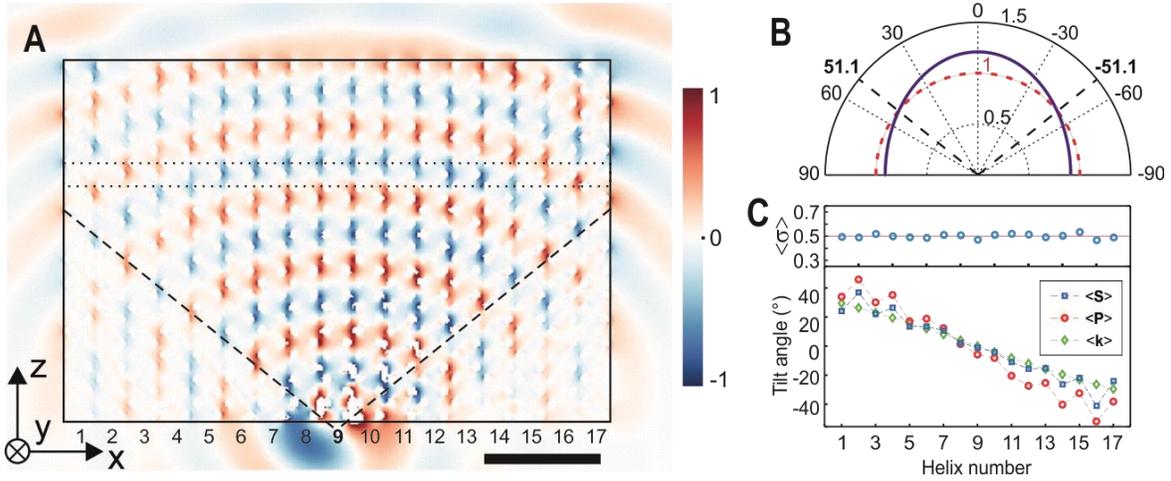

**Fig. 2. Calculated optical spin wave in a chain of homochiral helices.** A chain composed of seventeen 15-pitch helices is excited with a dipole source positioned at the lower end of the 9th nanostructure. The dipole is oriented along (0y). (**A**) Real part of the electric field component along (0x). Scale bar: 1.5 µm. (**B**) Angular distribution of the effective refractive index of the spin wave with the helix array. The center of the diagram coincides with the dipole position in the (0xz) plane. The dashed lines, tilted at ±51.1° (also visible in **A**), demonstrate the guided-to-leaky wave transition of the optical spin waves. (**C**) Dynamic properties of the optical spin wave averaged within the region delimited by the dotted lines. These lines are spaced by one period of the helix along (0z). Top: Helicity factor of the optical spin waves for each helix. Bottom: Tilt angle of the SAM and the wavevector and Poynting vector for each helix.

The elliptical distribution of the effective refractive index across the helix array (Fig. 2(B)) and the walk-off angle between the Poynting vector and wavevector (Fig. 2(C)) reveal the optical anisotropy of the helix structure. This anisotropy occurs due to the distinct phase velocities of the two orthogonal propagation axes in the lattice. The slightly asymmetric distribution of the optical waves propagating to either side of the excitation dipole is due to the asymmetric local energy coupling between the point-like source and the first helix.

Analogous to magnetic spin waves, optical spin waves are reflected at the interface separating the domains of different enantiomeric helices (see Fig. 3). On either side of the interface, the eigenmodes of the helices forming optical spin waves show orthogonal right or left circular polarizations (right and left-handed angular momenta). The blocking of the optical spin waves in the heterochiral lattice (see Fig. 3 (**B**)) is thus solely attributed to the angular momentum inversion at the boundary (see Supplementary text, Section S2). The optical field on the right side of the interface (Fig. 3 (**B**)) arises due to the interference between the incident and reflected optical spin waves, with some of the incident waves leaving the structure without interacting with the interface. The transmitted and reflected optical spin waves leaving the interface propagate at similar tilt angles with respect to the interface (Fig. 3(**C**)). The chirality-induced transmittance of the guided optical spin wave at the boundary between the enantiomeric domains is calculated to be less than 3.1%.

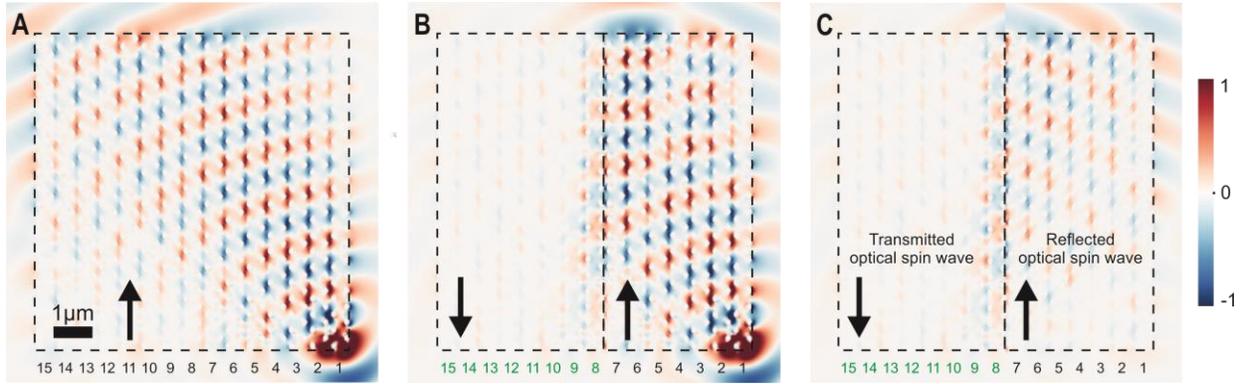

**Fig. 3. Reflection of a spin-wave at the interface between two homochiral helix chains with opposite handedness.** (**A**) Real part of the x-component of the electric field in a homochiral array composed of fifteen 19-pitch helices. The array is excited by a y-polarized dipole source positioned at the bottom end of the rightmost helix (λ = 1.55 μm). (**B**) The handedness of the final height helices is inverted, leading to two helix arrays with opposite handedness. In (**A**) and (**B**), the homochiral helix arrays are delimited with dashed lines. The helix handedness in each domain is represented with an arrow. (**C**) Transmitted and reflected optical spin waves on either side of the interface. The transmitted waves on the left side of the boundary correspond to raw data from simulations (see (**B**)), while the reflected field on the right side is obtained by subtracting the fields calculated within the homochiral array (cf. (**A**)) from the field calculated within the heterochiral array (cf. (**B**)).

A bidimensional helix array under normal light incidence is known to selectively block one of the two circular polarizations (one of the two available spin states of the incident waves) *(31)*. However, this spin-dependent reflection is lessened under oblique incidence conditions, as the intrinsic SAM of the impinging wave is tilted with respect to the helix axes (see Supplementary Fig. S4). This angular dependence is largely eliminated when using optical spin waves. For the conditions shown in Fig. 3(**B**), the transmittance across the heterochiral helix array remains below 3% for a range of tilted waves with incidence angles between 39° and 65°. Similar spin-dependent reflection has been investigated for spin waves at the interface between antiferromagnetically coupled ferromagnets separated by ultranarrow domain walls *(13-15)*.

### 4- Observation of optical spin waves

To experimentally test our concept of an optical spin wave, we studied energy transport within fabricated chains of helical nanoantennas (see the Materials and Methods section of *(24)*). Each of the ten elements constituting a chain consists of a six-turn gold-coated carbon helix lying on 100 nm cylindrical pedestals. The resulting helical nanoantennas were spaced 50 nm apart within the 1D lattices. Five chains of closely spaced nanohelices were used. The first chain includes only left-handed helical nanoantennas. The last four arrays include two successive domains of enantiomeric helices forming a heterochiral chain, with the first domain composed of left-handed helices. In these four last arrays, the number of left-handed helices forming the first domain varies from one to four. To evaluate energy transport within these fabricated helix chains, a local excitation at one end of the chain is applied with a rectangular nanoaperture at the pedestal of the rightmost helix.

Upon illumination of the substrate, the rectangular aperture acts as a background-free local excitation source for the helix array. Fig. 4 shows scanning electron micrographs of the fabricated helical structures. These images demonstrate helical antennas with limited surface roughness and minor geometrical discrepancies.

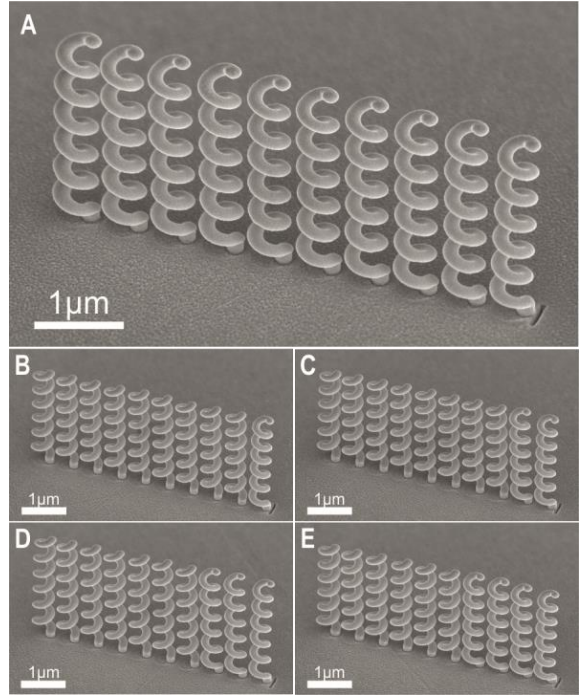

**Fig 4. Scanning electron microscopy images (oblique view) of helix arrays sustaining optical spin waves.** Each sample consists of ten 6-turn gold-coated carbon helices fabricated on a 100-nm thick gold layer deposited on a 1-mm-thick glass substrate. The helices are spaced 50 nm apart. The structures are locally excited with a rectangular nanoaperture set at the end of the rightmost helix. Upon back illumination of the substrate, the nanoaperture nonradiatively couples light to the helix chain as a dipole oriented parallel to the chain *(32)*. (**A**) Homochiral chain composed of left-handed helices. (**B**)-(**E**) Four heterochiral chains composed of two domains of enantiomeric helices: the first domain in contact with the rectangular nanoaperture is composed of one to four left-handed helices.

Figure 5 shows a comparison of the measured and calculated far-field diffraction patterns for illumination from the substrate with linearly polarized light at a wavelength of 1570 nm. Since the optical waves leaving the chain hardly deviate from the optical spin waves (see Fig. 3(A)), the far-field pattern allows us to investigate energy transport within the structure. We measured the far-field angular diffraction pattern by imaging the helix chain with a specific optical bench that projects the Fourier plane (i.e., the back focal plane) of a microscope objective focused on the sample onto an infrared camera (see the Materials and Methods section of *(24)*). The camera data are then converted into angular emission distributions *(33)*.

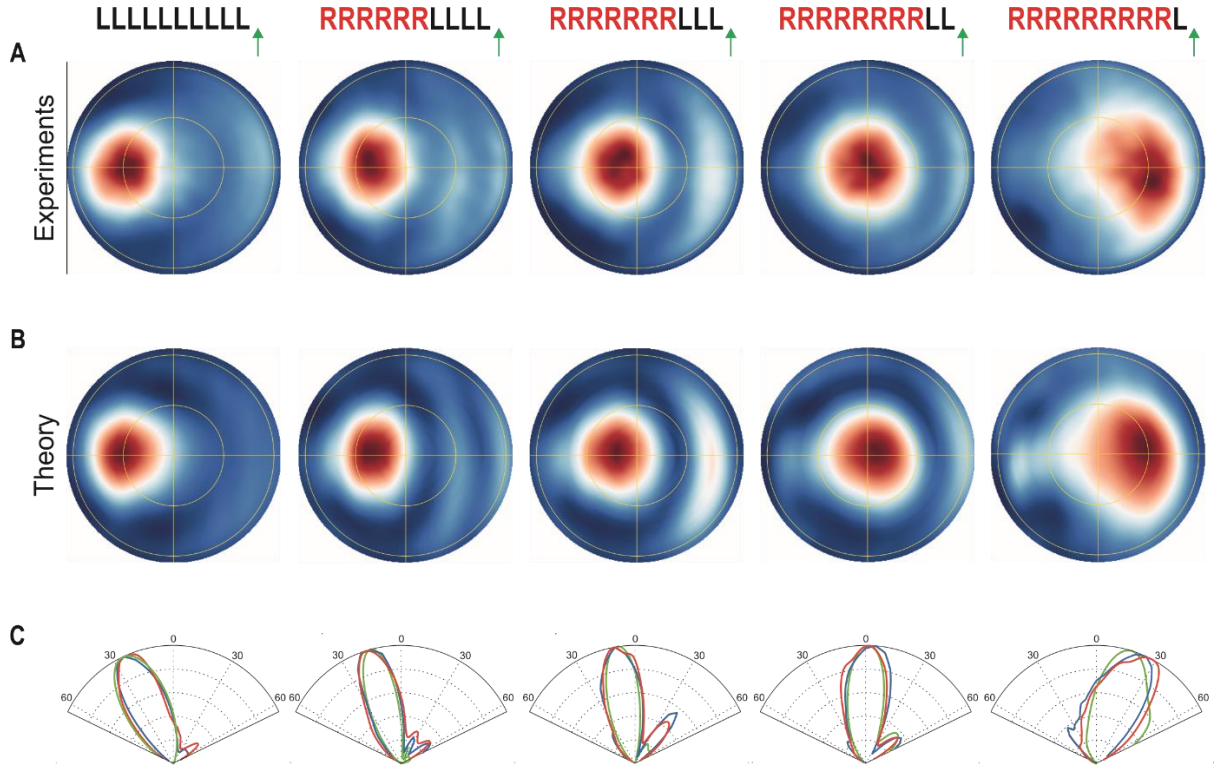

**Fig. 5**. **Far-field measurement of optical spin waves**. Normalized experimental diffraction patterns (row (**A**)) and numerically calculated theoretical patterns (row (**B**)) for the five configurations of the helix arrays shown in Fig. 4. Each column in Fig. 5 shows the distribution of the helices depicted above each pattern in row (**A**). The letters R and L denote right- and left-handed helices, respectively, while the green arrows indicate local dipole excitation of the chiral chains on their right edge. The crosscuts through the far-field patterns in rows (**A**) and (**B**) along the direction of the helix chains are plotted in row (**C**) for each helix array. The theoretical and measured diffraction profiles are plotted in blue and red, respectively. These zenithal angular plots are compared to those calculated with a simplified model, in which the optical spin wave and the heterojunction are represented with a tilted Gaussian beam and a gold mirror, respectively (green profiles; see Supplementary material).

The measured far-field patterns shown in Fig. 5(**A**) reveal the theoretically anticipated chirality-dependent blocking of the optical spin waves. When all the helices have the same handedness, the optical spin waves are transmitted through the structure (first column in Fig. 5). When the structure is locally excited on the right side, the emanating light waves are tilted to the left, confirming that optical spin waves travel from right to left within the helix array. When the second and subsequent helices are inverted (fifth column in Fig. 5), the spin waves are back-reflected, as the optical waves are mainly tilted to the right, with a tilt angle opposite to that of the transmitted waves within the homochiral array (first column in Fig. 5). Considering the limited length of the fabricated helices, this extreme configuration guarantees that all the guided waves interact with the heterochiral interface. When the first inverted helix is moved from the third to the fifth position within the chain, some of the optical spin waves leave the array without interacting with the inverted domain. In these intermediate configurations, the optical waves leaving the helix array are tilted toward various directions between those observed in the two extreme configurations (see the second, third, and fourth columns in Fig. 5(**A**)). The numerical predictions shown in Fig. 5(**B**) agree well with the experimental findings. A comparative analysis was performed with calculations based on a tilted Gaussian wave reflected by a gold mirror, indicating that the heterochiral interface acts as a reflector of optical spin waves (Fig. 5(**C**)). The details of the calculation model are presented in Fig. S6 of *(24)*.

### 5- Discussion

We show that one-dimensional arrays of coupled plasmonic nanohelices enable electromagnetic energy transport in the form of optical waves with angular momentum determined by the helix handedness. The present light waves have SAM determined by the chiral properties of matter and photonic spin-orbit interaction. These optical waves share similar properties with spin waves associated with collective electron spin excitations in ferromagnets and can be considered their photonic analog. Quantization of optical spin waves may lead to optical magnons based on the collective excitation of chiral plasmon nanoparticles.

The abilities of helical antennas to produce optical spin waves at any frequency from visible light to microwaves and control the frequencies of these traveling waves irrespective of the material permittivity may lead to new research directions in new optical effects, optical circuits and functionalities over a broad spectral range for information processing and transfer and molecular sensing. Moreover, other forms of optical spin waves may exist within lattices composed of various chiral particles, demonstrating the potential scope of exploration within this field.

Optical spin waves can be guided in chains composed of subwavelength structures and reflected by individual elements with antiparallel SAM, which is promising for SAM transfer and manipulation in compact integrated platforms. Spin-optical functionalities may arise by combining diverse operations with optical spin waves, including waveguiding, focusing, interference, steering and modulation, within metamaterials and metasurfaces, possibly merging positive and negative refractive indices.

**Acknowledgments:** The authors thank Patrice Genevet for fruitful discussions. This research was supported by the French Agency of Research (contract ANR-18-CE42-0016), the Region "Bourgogne Franche-Comte" and the EIPHI Graduate School (contract ANR-17-EURE-0002). This work was also supported by the French RENATECH network and its FEMTO-ST technological facility.

# Supplementary Materials

**Materials and Methods**

Fabrication of the helix chains

First, 1D arrays of carbon helices were realized by growing ten 105-nm large carbon wires on a 100-nm thick gold layer using focused ion beam-induced deposition (FIBID) *(34,35)*. Then, the overall structures were plasma sputtered with a thin gold film using glancing angle deposition *(32,36,37)*. Gold was chosen for its high conductivity at near-infrared wavelengths and its resistance to oxidation. During gold deposition, the metal target was tilted by an angle of 80° from the helix axis, and the sample was rotated at a constant speed of 2 rev min$^{-1}$. With a pitch length of 370 nm and an outer diameter of 505 nm, the fabricated helices were slightly scaled down with respect to the anticipated designs to compensate for the spectral redshift of their response regarding expectations. This spectral redshift is due to a nonhomogeneous gold coating of the 3D structures due to a shadow effect in the metal deposition *(32,38)*. To probe energy transport within these fabricated chains of helices, a local excitation at one extremity of the chain is realized by milling with a FIB a rectangular nanoaperture right at the pedestal of the rightmost helix.

Detection system

We used the optical bench represented in Fig. S5. A 1570 nm continuous wave laser beam emerging from a fibered tunable laser (EXFO-T100S) was collimated and focused with a microscope objective (Olympus LCPlan N 50x, 0.65 IR) to back illuminate the sample (from the glass substrate). The objective was mounted onto an inverted microscope (NIKON, TE2000). The sample was attached onto a high-resolution 3D piezo-stage (Mad City Labs) also mounted onto the inverted microscope. A half-wave plate (AHWP05M-1600, Thorlabs) and a linear polarizer (LPNIR100-MP2) were positioned before the microscope objective to create a linear polarization parallel to the short side of the rectangular nanoaperture, thereby ensuring optimum light coupling to the helix arrays. Light transmitted through the sample was detected with an infrared camera (GoldEye model G-033, Allied Vision) coupled to a high numerical aperture objective (Leitz Wetzlar NPL FL 63x, 0.9). To image the far-field pattern of the structures, two lenses (Lens 1 and Lens 2, see Fig. S5) were inserted between the top objective and the camera to project the back focal plane of the top objective onto the camera.

Data Processing

The far-field patterns acquired with our optical bench were numerically separated from the noise and few imperfections with a simple digital low-pass filtering process. We selected a cutoff spatial frequency for the filter that was sufficiently high to not affect the far-field patterns. Imperfections were composed of high-frequency interference fringes arising from spurious multiple reflections at the air-glass interfaces of our dioptric system.

**Supplementary Text**

S1. Dispersion relation of the eigenmodes of a plasmonic helix.

The dispersion relation of the twisted modes of the helix (Fig. S1) is derived through a series of finite difference time domain (FDTD) simulations at various wavelengths. These simulations involve the calculation of the electric field along a 32-turn helix, which is excited with a dipole source positioned at one of its ends (as shown in the right panel of Fig. S1). The specific parameters of the helical structure, including the pitch length, outer diameter, and wire diameter, correspond to those envisioned for the intended design. The dipolar source is aligned along the (0x)-axis. In each FDTD simulation, the y-component of the electric field (perpendicular to the dipole) is recorded in the (0yz)-plane, which contains the helix axis. To determine the real part of the wavevectors associated with the guided and leaky waves along the helix, the electric field along the structure is Fourier transformed. In this transformation, an analytical formula is used to address potential resolution limitations inherent to fast Fourier transform processes. The dispersion relation of the modes is determined by accumulating the optical eigenvalues of the helix calculated for each wavelength and applying an interpolation method to establish comprehensive dispersion curves.

S2. Chirality-induced reflection of an optical spin wave.

To ensure that the reflection of the optical spin waves at the interface between two helix domains of opposite handedness is influenced by only the chirality (and not by any alterations in interhelix coupling), we conducted simulations involving both homochiral and heterochiral helix chains (see Fig. S3 (**A**) and (**B**), respectively). In these simulations, the $8^{th}$ helix from the right was rotated along its axis (see the figure insets). In the heterochiral array (Fig. S3 (**B**)), the $8^{th}$ helix is the first inverted structure in the second domain. As depicted in Fig. S3, the propagation and reflection mechanisms of optical spin waves are not affected by local modifications due to interhelix coupling resulting from the rotation of a single structure. This finding demonstrates the chirality-induced reflection of the optical spin waves.

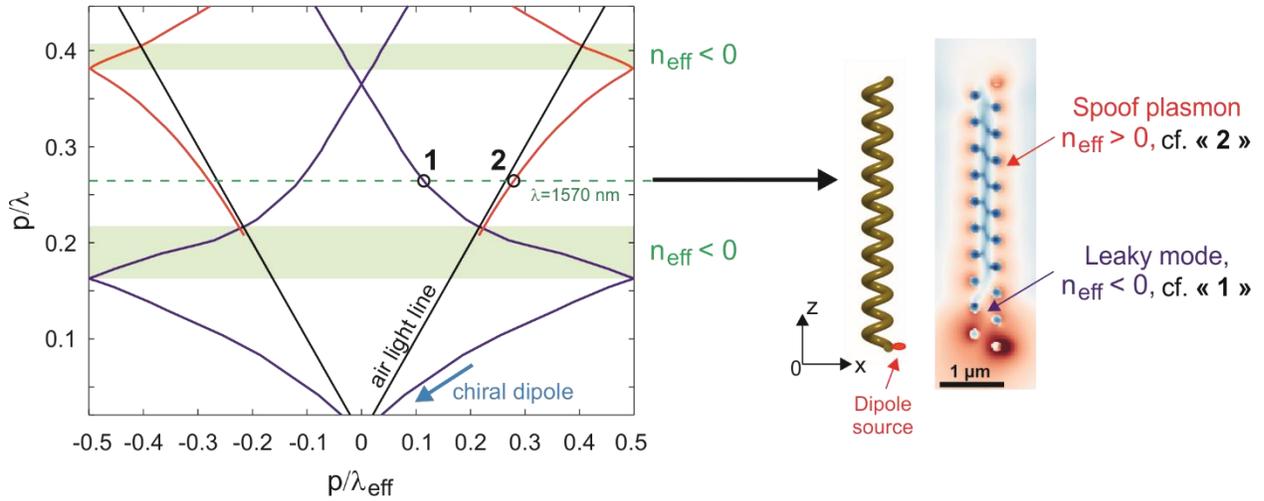

**Fig. S1.**
Guided and leaky modes within a plasmonic nanohelix. Left: Calculated dispersion relation of the helix eigenmodes (real part of the effective wavelength versus the vacuum wavelength). The green regions denote the spectral domains in which one of the guided modes has a negative effective index. Right: Calculated cross-section of the amplitude of the optical electric field along a 10-turn helix excited with a dipolar source at $\lambda=1570$ nm (the dipole is oriented along (0x)). The guided and leaky modes of the structure are identified with the numbers "2" and "1", respectively, in the field distribution and the dispersion curves.

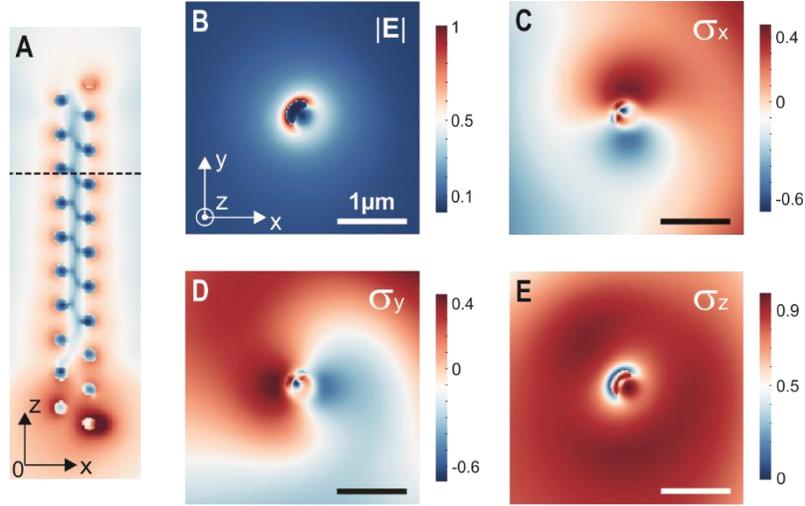

**Fig. S2.**

Spin vector texture within a transverse cross-section of a 10-turn helix. (**A**) Longitudinal cross-section of the calculated optical electric field (amplitude) along a 10-turn helix excited with a dipolar source at a wavelength of 1570 nm. The dipole is x-polarized and located as indicated in Fig. S1. (**B**) Calculated amplitude of the optical electric field in the transverse cross-section of the helix indicated by the dashed line in (**A**). (**C**)-(**E**): Calculated local polarization helicity factors along (0x), (0y) and (0z) within the same transverse plane ($\sigma_x$, $\sigma_y$ and $\sigma_z$, respectively). The guided spoof surface plasmons show a 3D spin texture approaching that of a Bloch skyrmion *(11)*. The local polarization helicity, described by Eq. 2, obtains its maximum value along the helix axis, with the value approaching 1.

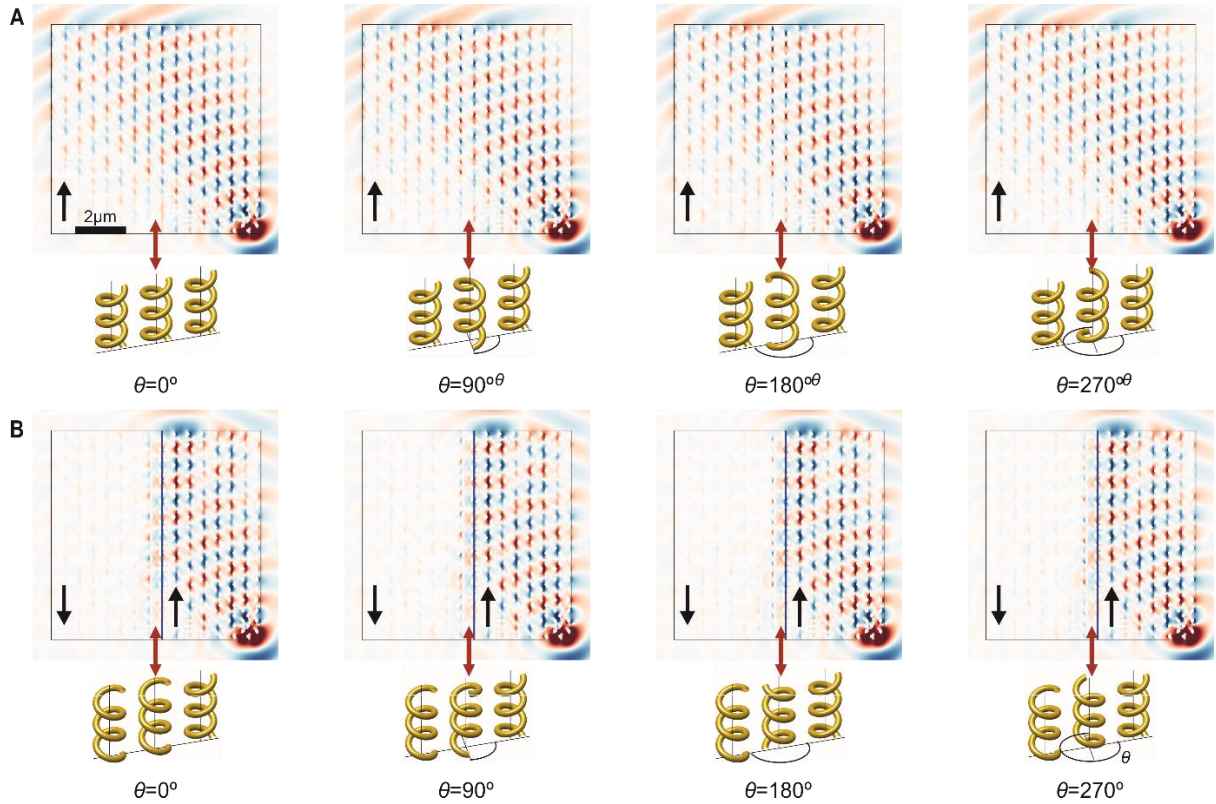

**Fig. S3.**
The reflection of an optical spin wave at the interface between two domains of enantiomeric helices occurs due to chirality. (**A**): Calculated x-component of the optical electric field across a homochiral array consisting of fifteen 19-turn helices, considering four distinct orientations of the 8th helix (the rotation angle of the 8th helix is illustrated in the figure insets). The array is excited with a y-polarized dipole (out-of-plane dipole) in contact with the lower end of the rightmost helix. (**B**) Same configuration as the heterochiral chain: the 8th helix and the following structures are inverted, forming two domains of enantiomeric helices.

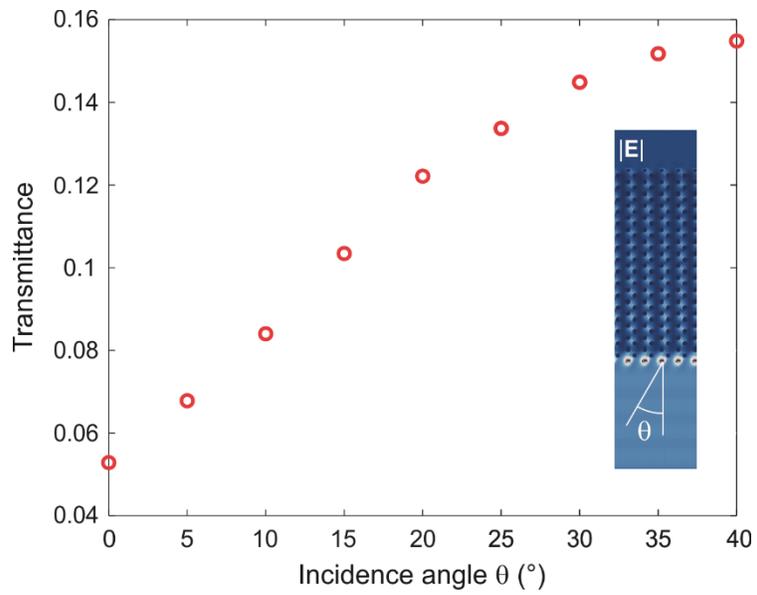

**Fig. S4.**
Transmittance of a plane wave through an infinite square lattice composed of 16-turn helices as a function of the incidence angle θ (see inset). With normal incidence, the helix structure blocks the incoming circularly polarized wave, resulting in a transmittance of 5.3%. The transmittance increases significantly by three times as the incidence angle increases by 40°.

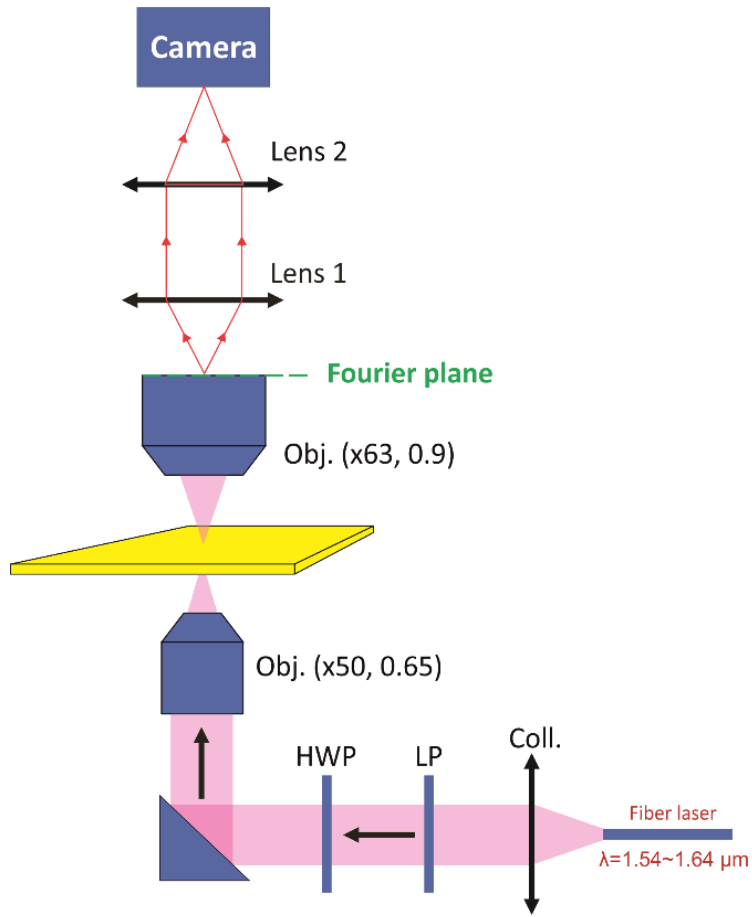

**Fig. S5.**
Schematic diagram of the experimental setup used to probe optical spin waves. HWP: half-wave plate, LP: linear polarizer, Coll. : fiber collimator.

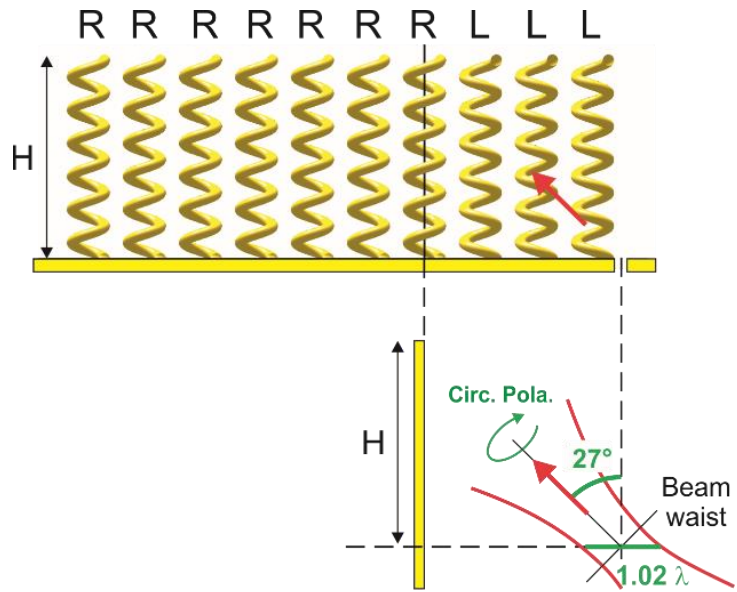

**Fig. S6.**
Simplified model of the experimentally studied heterochiral helix array. In this model, the interface between the enantiomeric helix domains is represented by a gold mirror positioned at the axis of the first flipped helix in the second domain. The position of the mirror is determined according to the field distribution shown in Fig. 3(**B**), showing a domain wall covering the first inverted helix. The optical spin wave is modeled with a tilted Gaussian beam with a carefully chosen tilt angle and width (at the waist) to closely approximate the propagation characteristics of the optical spin wave, as represented in Fig. 3(**A**). The center of the beam waist coincides with the center of the rectangular nanoaperture used to experimentally excite the helix array. The calculations are performed with the 2D FDTD method.